\documentstyle[aps,prl,epsf,floats,twocolumn]{revtex}
\begin{document}
\draft
\twocolumn[\hsize\textwidth\columnwidth\hsize\csname @twocolumnfalse\endcsname
\title{Effects of a Parallel Magnetic Field on the Metal-Insulator Transition 
in a Dilute Two-Dimensional Electron System}
\author{Kevin Eng$^{1,2}$, X. G. Feng$^{1,*}$, Dragana Popovi\'{c}$^{1}$, 
and S. Washburn$^{2}$}
\address{$^{1}$National High Magnetic Field Laboratory, 
Florida State University, Tallahassee, FL 32310 \\ 
$^{2}$Department of Physics and 
Astronomy, The University of North Carolina at Chapel Hill, Chapel Hill, NC 
27599}
\date{\today}
\maketitle

\begin{abstract}

The temperature dependence of conductivity $\sigma (T)$ of a two-dimensional
electron system in silicon has been studied in parallel magnetic fields 
$B$.  At $B=0$, the system displays a metal-insulator transition at a critical
electron density $n_c(0)$, and $d\sigma/dT >0$ in the metallic phase.  At low 
fields ($B\lesssim 2$~T), $n_c$ increases as $n_c(B) - n_c(0) 
\propto B^{\beta}$ ($\beta\sim 1$), and the zero-temperature 
conductivity scales as 
$\sigma (n_s,B,T=0)/\sigma (n_s,0,0)=f(B^{\beta}/\delta_n)$ 
(where $\delta_n=(n_s-n_c(0))/n_c(0)$, and $n_s$ is electron density) as 
expected for a quantum phase transition.  The metallic phase persists in 
fields of up to 18~T, consistent with the saturation of $n_c$ at high fields.
\end{abstract}

%
%
%
%
]

The possibility of a metal-insulator transition (MIT) in two dimensions (2D) 
has been a subject of intensive research in recent 
years~\cite{Ab_review,novel}.  Most studies have focused on the conventional 
metallic behavior where the conductivity $\sigma$ increases with decreasing 
temperature $T$ ({\it i.~e.} $d\sigma/dT<0$) but the origin of such 
$\sigma (T)$ is still not understood.  Similarly, a dramatic decrease of 
$\sigma$ with magnetic fields $B$ applied parallel to the 2D plane remains
puzzling.  Above some characteristic field, which is a function of the carrier 
density $n_s$, this unexpectedly large negative magnetoconductance is 
followed by a weaker dependence on $B$.  Parallel $B$ has been also shown to
suppress the metallic $T$-dependence 
($d\sigma/dT<0$)~\cite{Btempdep,okamoto,yoon} and transform it into an 
insulatinglike behavior ($d\sigma/dT>0$).  Since it has been commonly assumed 
that $d\sigma/dT>0$ is necessarily related to an insulating state at 
$T=0$, this was taken as evidence that high parallel $B$ eliminates the MIT and
the 2D metallic phase entirely~\cite{Ab_review}.  The field where 
$d\sigma/dT$ changes sign (for a fixed $n_s$) has been identified as the 
critical field for a field-induced MIT~\cite{yoon}.  Furthermore, it has been 
established that this critical field is comparable both to the field
where the crossover from a low-field to high-field regime takes place and to 
the field where 2D carriers become fully 
spin-polarized~\cite{okamoto,spinpol_vit,spinpol,beta_Pudalov}, suggesting 
that the 2D metallic phase may exist only in a spin-unpolarized system.  By 
contrast, the methods that do not use the sign change of $d\sigma/dT$ as a 
criterion to determine $n_c$ have yielded~\cite{beta} finite values of $n_c$ 
even in high parallel $B$ where $d\sigma/dT>0$ at all $n_s$, which implies
that the 2D metal may exist even for spinless electrons.  It is 
clear that the fate of the metallic phase in a parallel $B$ represents 
one of the major open issues in the studies of dilute, strongly interacting
2D systems.

The main difficulty associated with the attempts to resolve this issue so 
far~\cite{Btempdep,okamoto,yoon,spinpol_vit,spinpol,beta_Pudalov,beta} has 
been that $\sigma (T)$ does not have a simple form for 
$n_s>n_c(0)$ in parallel $B$.  Therefore, this does not allow one to make 
reliable extrapolations to $T=0$ and to establish the existence of a true MIT 
with high credibility.  In addition, there is no theoretical 
justification~\cite{scale} for assuming that the 2D metallic phase can be 
characterized only by $d\sigma/dT<0$.  In disordered 3D metals, for example, 
it is well known that this derivative can be either negative or positive near 
the MIT~\cite{Lee}.  More relevant to this work, a recent study~\cite{novel} 
of a 2D electron system in Si metal-oxide-semiconductor field-effect 
transistors (MOSFETs) in $B=0$ has established unambiguously that the metallic
phase may be described by $d\sigma/dT>0$ such that $\sigma$ decreases to a 
nonzero value as $T\rightarrow 0$.  This occurs when some of the localized 
states in the tail of the upper electric subband~\cite{tails} also become 
populated~\cite{DP_MIT,spinflip} as the subband splitting is reduced by 
applying voltage ($V_{sub}$) to the Si substrate~\cite{AFS}.  Such localized 
states act as additional scattering centers for 2D electrons~\cite{subbands}, 
and perhaps may even act as local magnetic moments if they are singly occupied
(due to a large on-site Coulomb repulsion)~\cite{spinflip,moments_comment}.  

Regardless of the detailed microscopic picture, the 2D metal with 
$d\sigma/dT>0$ in $B=0$ exhibits two features~\cite{novel} that distinguish it
from the $d\sigma/dT<0$ case, and that are of crucial importance for the study
of $\sigma$ in parallel $B$.  First, $\sigma (T)$ follows a simple 
and precise form over a very broad (two decades) range of $T$: 
$\sigma (n_s,T)=\sigma (n_s,T=0)+C(n_s)T^2$.  This allows a reliable 
extrapolation to $T=0$, which yields $\sigma (n_s,T=0)\sim\delta_{n}^{\mu}$
($\mu\approx 3$), as expected in the vicinity of the 
MIT~\cite{scale,Goldenfeld}.  Second, near the MIT, the data obey dynamical 
scaling $\sigma (n_s,T)=\sigma_c(T)f(T/\delta_{n}^{z\nu})$ with 
$\sigma_c=\sigma(n_s=n_c,T)\sim T^x$ ($z\nu=1.3\pm 0.1$, $x\approx 2.6$, 
$\mu=x(z\nu)=3.4\pm0.4$), both in agreement with theoretical expectations near
a quantum phase transition~\cite{scale} and consistent with the extrapolations
of $\sigma (T)$ to $T=0$.  Here we show that such features allow
us to establish a phase diagram in the $(\delta_n,B,T=0)$ plane.  We find that
the behavior of low-field $\sigma (B)$ is consistent with the phase diagram, 
and can be attributed to an increase in $n_c(B)$, as expected near a 
true MIT~\cite{scale}.  The saturation of $n_c(B)$ observed at higher $B$ 
strongly suggests that the 2D metal may exist even for spinless electrons.
 
Measurements were carried out on n-channel Corbino shaped Si MOSFETs (channel
length $=0.4$~mm, mean circumference $=8$~mm) with poly-Si gates, self-aligned
ion-implanted contacts, and the peak mobility 
$\sim 1$~m$^2$/Vs at 4.2~K.  Other sample details have been given 
elsewhere~\cite{DP_MIT,spinflip,novel}.  $n_s$ was controlled by the gate
voltage $V_g$.  $V_{sub}=+1$~V was applied to minimize the subband splitting
and maximize the $T$-range (up to $\sim 2$~K) in which the metallic behavior 
with $d\sigma/dT>0$ is observed~\cite{spinflip,novel}.  
Conductance was measured as a function of $V_g$ using standard low-noise 
analog lock-in techniques at $\sim 17$~Hz and a low-noise current preamplifier
(the lead resistance was subtracted by the usual method).  Excitation voltages
were kept low enough to ensure that the conduction was Ohmic at the lowest $T$
(see Ref.~[2] for more details).
A systematic study of $\sigma (n_s,T)$ was done in a He$^3$ cryostat in static
parallel fields up to 9~T and for $0.25\leq T\leq 4$~K.  
Measurements up to 18~T were carried out with the sample in the mixing chamber
of a dilution refrigerator with a base $T\approx 0.020$~K.  

\begin{figure}
\epsfxsize=3.2in \epsfbox{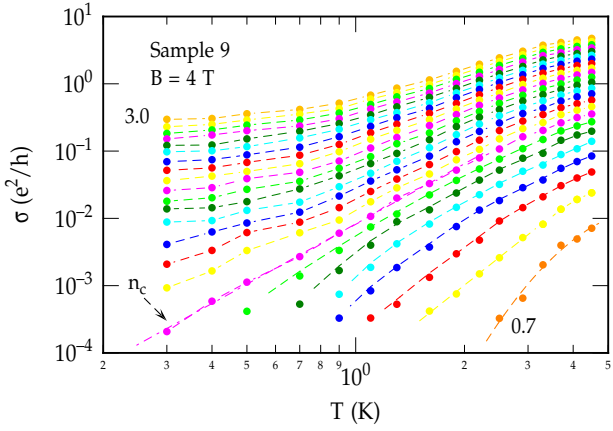}\vspace{5pt}
\caption{$\sigma (T)$ for sample 9 at $B=4$~T.  $n_s$ varies from 
$3.0\times 10^{11}$cm$^{-2}$ (top) to $0.7\times 10^{11}$cm$^{-2}$ (bottom) in
steps of $0.1\times 10^{11}$cm$^{-2}$.  $n_c(B=4$~T$)=1.4\times 
10^{-11}$cm$^{-2}$ and is marked by the arrow.  The dashed lines guide the 
eye.  $\sigma_c$ clearly follows a simple power-law dependence on $T:\sigma_c
\propto T^x$.
\label{sigma}}
\end{figure}

Figure~\ref{sigma} shows $\sigma (T)$ for sample 9 (also studied in Ref.~[2]) 
at $B=4$~T for various $n_s$ in the vicinity of the MIT.  The data are
qualitatively similar to the $B=0$ case~\cite{novel}.  For example, at
the lowest $n_s$ ({\it i.~e.} $n_s<n_c$), $\sigma$ decreases exponentially with
decreasing $T$, indicating an insulating state at $T=0$.  In the metallic
phase ($n_s>n_c$), $\sigma (T)$ is weaker and its {\em curvature} is the 
opposite from the one expected for an insulating state.  It clearly 
extrapolates to a finite value as $T\rightarrow 0$, as discussed in more 
detail below.  As shown in Fig.~\ref{sigma}, $n_c$ is identified  as the 
density where $\sigma_c =\sigma(n_s=n_c,T) \propto T^x$, consistent with the 
$B=0$ case and in agreement with general arguments~\cite{scale}.  The 
exponent $x$ is found to be $2.7\pm 0.4$, and remains constant as a function 
of $B$.  

\begin{figure}
\epsfxsize=3.2in \epsfbox{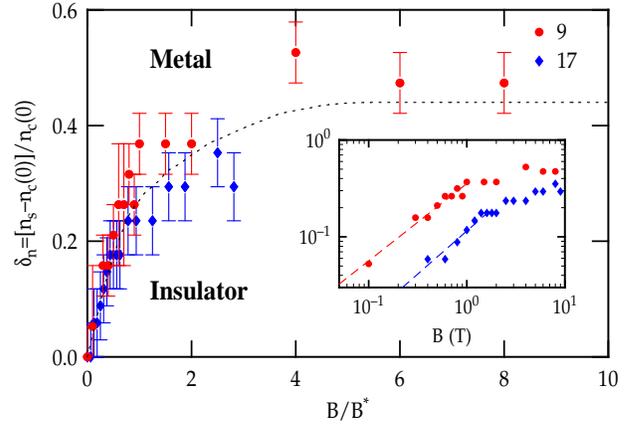}\vspace{5pt}
\caption{$T=0$ phase diagram for two samples.  The dashed line guides the eye.
The boundary between metallic and insulating phases is described by a 
power-law relation (see inset) $[n_c(B)-n_c(0)]/n_c(0)\propto 
(B/B^{\ast})^{\beta}$ at low fields, with the same crossover exponent 
$\beta\approx 0.9$ for both samples ($B^{\ast}=1$~T for sample 9).  Inset: the 
same data {\it vs.} $B$ on a log-log scale.  The dashed lines are fits with 
the slopes equal to $\beta$.  At $B=0$, $n_c(10^{11}$cm$^{-2})=0.95\pm 0.05$ 
and $0.85\pm 0.05$ for samples 9 and 17, respectively.  
\label{phase}}
\end{figure}

The critical density $n_c$ was determined in this way at each given $B$.  
Fig.~\ref{phase} inset shows the relative change of $n_c(B)$ with respect to 
its zero-field value
$n_c(0)$ as a function of $B$ for two different samples.  At low fields 
($B<2$~T), $n_c(B)$ increases with $B$ in a power-law fashion:
$[n_c(B) - n_c(0)]/n_c(0)\propto B^{\beta} $, where the crossover exponent 
$\beta=0.80\pm 0.06$ and $0.9\pm 0.1$ for samples $9$ and $17$, respectively. 
We point out that such a power-law shift of $n_c$ with $B$ is 
expected to occur in the case of a true MIT~\cite{scale}, and has been 
observed in several 3D systems~\cite{3Dbeta}.  In the case of conventional 2D 
metallic behavior in high-mobility Si MOSFETs, 
finite values of $n_c(B)$ were found based on vanishing activation energy and 
vanishing nonlinearity of current-voltage curves as extrapolated from the
insulating side~\cite{beta}.  For those data,
we find that the corresponding $\beta=1.1\pm 0.1$.  It is interesting that the
crossover exponents $\beta$ are almost the same in both cases (Ref.~\cite{beta}
and Fig.~\ref{phase}) even though the $B=0$ behaviors of $d\sigma/dT$ are very 
different.  Furthermore, in both cases the saturation of $n_c(B)$ is observed 
at higher fields.  The determination of $n_c(B)$ based on a sign change of 
$d\sigma/dT$ has yielded $\beta\approx 1$ in high-mobility Si 
MOSFETs~\cite{beta_Pudalov}, and $\beta\approx 1.4$ in a 2D hole system in 
GaAs~\cite{yoon}.  While the difference in $\beta$ between Si and GaAs devices
might be due to the difference in the dominant scattering mechanisms in the 
two materials, it should be noted that the mere sign of $d\sigma/dT$ is 
clearly not a reliable indicator of the nature of the $(T=0)$ ground state.
 
Since $\beta\approx 0.9$ for both of our samples and only the absolute values 
of $n_c(B)$ are different, it is possible to create a generalized phase 
diagram in the $(\delta_n,B,T=0)$ plane as shown in Fig.~\ref{phase}.  
The dashed line in Fig.~\ref{phase} represents the boundary between metallic 
and insulating phases.  It is clear that, for a fixed, small $\delta_n$, the 
system will undergo a magnetic field-driven MIT, leading to a strong 
suppression of magnetoconductance (MC).  On the other hand, for densities far 
from $n_c(0)$ ({\it i.~e.} large $\delta_n$), the 2D system will remain 
metallic with increasing $B$.  In that case, one expects a much weaker MC.  
Exactly this kind of behavior of MC has been observed in both our 
samples~\cite{eng_conf} and high-mobility Si MOSFETs in the conventional 
metallic regime~\cite{spinpol_vit}.

\begin{figure}[t]
\epsfxsize=3.2in \epsfbox{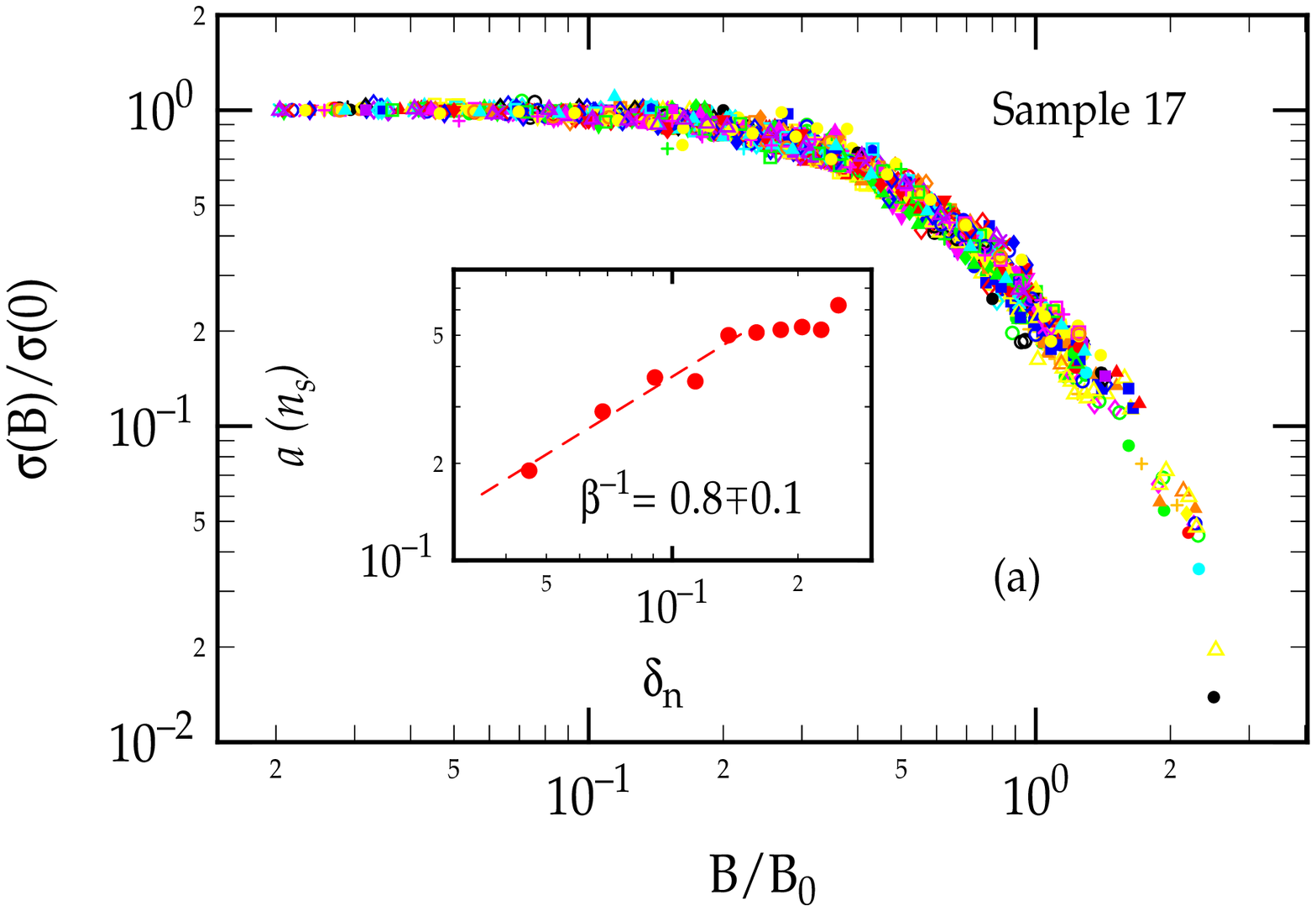}\vspace{5pt}
\epsfxsize=3.2in \epsfbox{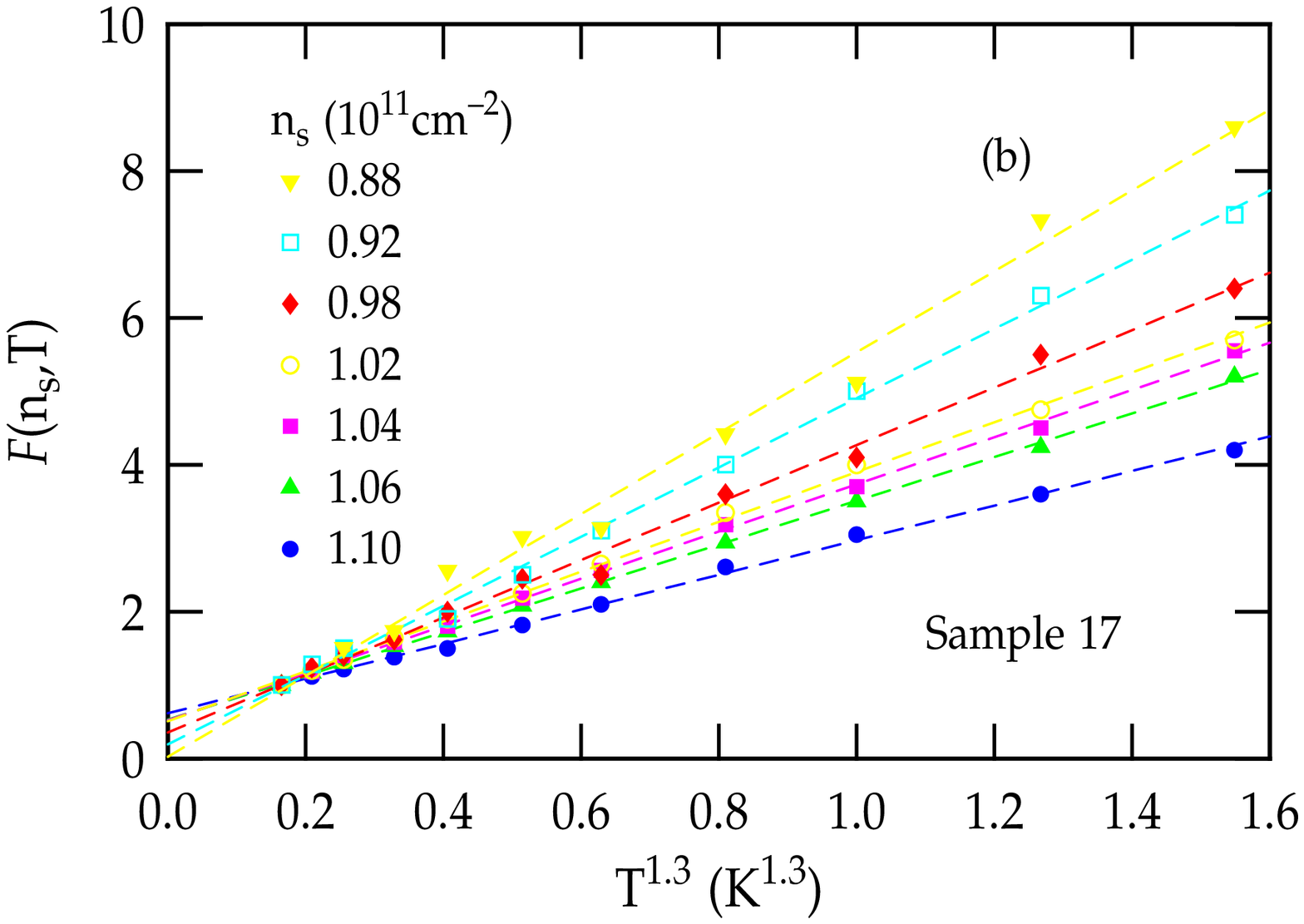}\vspace{5pt}	
\caption{(a) Scaling of low-field ($B\protect\lesssim 2$~T) 
magnetoconductivity for
$n_s(10^{11}$cm$^{-2})=0.80-1.10$ in steps of 0.02, and ten different $T$ from
0.25 to 1.4~K.  The scaling parameter $B_0(n_s,T)={\cal A}(n_s){\cal 
F}(n_s,T)$.  (b) ${\cal F}(n_s,T)=a(n_s)+b(n_s)T^{1.3}$, as shown for several 
$n_s$.  $a(n_s)\propto \delta_{n}^{1/\beta}$, as shown in the inset of (a).
\label{mcscaling}}
\end{figure}

First we focus on the regime of low fields ($B\lesssim 2$~T) and small 
$\delta_n$, where MC exhibits a large drop~\cite{eng_conf}.  We find that all
the MC curves $\sigma(B)/\sigma(0)$ for a fixed $n_s$ and different $T$ can be
collapsed onto one function using a single scaling parameter 
${\cal F}(n_s,T)$ [Fig.~\ref{mcscaling}(a)].  For each given $n_s$, we find 
empirically that ${\cal F}(n_s,T)=a(n_s)+b(n_s)T^{1.3\pm 0.2}$ 
[Fig.~\ref{mcscaling}(b)].  The parameter $a(n_s)$ decreases with 
decreasing $n_s$ and extrapolates to zero at a density equal to $(0.88\pm 
0.04)\times 10^{11}$cm$^{-2}$.  Within our measurement error, this density is 
the same as $n_c(0)$.  Fig.~\ref{mcscaling}(a) inset 
shows that a fit to the critical form, $a(n_s)=a_{0}\delta_{n}^{1/\beta}$, 
yields a crossover exponent $\beta=1.2\pm 0.2$ ($a_0=1.7\pm 0.4$), consistent
with the value of $\beta$ that was established independently earlier 
(Fig.~\ref{phase}) based on the extrapolations of $\sigma (T)$ to $T=0$ in the
metallic regime at fixed $B$ (Fig.~\ref{sigma}).  
Moreover, Fig.~\ref{mcscaling}(a) shows that it is
also possible to collapse $\sigma (B)/\sigma(0)$ for all different $n_s$ and 
$T$ using a single scaling parameter $B_0(n_s,T)={\cal A}(n_s){\cal 
F}(n_s,T)$, which combines the previous scaling parameter ${\cal F}$ with a
density dependent prefactor ${\cal A}(n_s)$.  The fitting parameters 
${\cal A}(n_s)=(2.8\pm 0.3)n_s-(1.6\pm 0.3)$ and 
$b(n_s)=-(11.4\pm 0.9)n_s+(14.9\pm 0.9)$ (all $n_s$ in units of 
$10^{11}$cm$^{-2}$) do not show much
variation near $n_c(0)$.  

Therefore, our results demonstrate that the 
zero-temperature conductivity scales with $n_s$ and $B$ as
\begin{equation}
\sigma(\delta_n,B,0)/\sigma(\delta_n,0,0)=f(B/\delta_{n}^{1/\beta}),
\end{equation}
where $\beta\sim 1$.  For a fixed $n_s$, the crossover function 
$f(B/\delta_{n}^{1/\beta})$ decreases by orders of magnitude with increasing 
$B$ [Fig.~\ref{mcscaling}(a)] as the system approaches the magnetic 
field-driven MIT.  The critical fields $B_c$ for such a transition are given 
by $B_c\propto\delta_{n}^{1/\beta}$ in agreement with the phase diagram 
determined earlier (Fig.~\ref{phase}).  It is clear that the scaling parameter
$B_0(n_s,T=0)\propto B_c$.  We also note that the scaling plot in 
Fig.~\ref{mcscaling}(a) includes a few densities below $n_c(0)$ in the 
quantum critical region, where ${\cal F}(n_s,T)\propto T^{1.3}$ in the given 
$T$ range.  Furthermore, for $n_s=n_c(0)$, our results show that  
$\sigma(0,B,T)/\sigma(0,0,T)=f(B/T^{1.3})$.  According to standard scaling 
arguments~\cite{scale}, the exponent $1.3=\beta/z\nu$.  With $\beta=1.2\pm 
0.2$, this yields $z\nu=0.9\pm 0.3$, in reasonable agreement with $z\nu=1.3\pm
0.1$ determined from scaling of $\sigma(n_s,T)$ in $B=0$~\cite{novel}.  All 
these findings provide strong experimental evidence for a quantum phase 
transition in {\em both} $B=0$ and $B\neq 0$ in our system.  

Recent studies of some Si MOSFETs in the conventional metallic regime
have shown~\cite{city_scaling} that, in that case, the MC curves can be
also scaled with a single parameter $B_{\sigma}$, where 
$B_{\sigma}$ corresponds to the onset of full spin polarization of the 
electron system~\cite{sergey_scaling} ($\beta\approx 1.7$ and $\beta\approx 1$
in Refs.~\cite{city_scaling} and \cite{sergey_scaling}, respectively).  
However, since 
the form of $\sigma(T)$ in the conventional metallic regime does not allow 
reliable extrapolations to $T=0$ in $B\neq 0$, it is difficult to establish 
the existence of a metallic phase and $n_c(B)$ (or, equivalently, $B_c(n_s)$) 
with certainty.  We showed earlier~\cite{DP_MIT,spinflip} that our Si MOSFETs 
exhibit exactly the same kind of conventional metallic behavior in absence of 
scattering by 
local magnetic moments.  Therefore, we expect that similar values of $B$ will
be required for full spin polarization of our 2D system at comparable $n_s$,
even when a small fraction of electrons occupies the tail of the upper subband.
Our work demonstrates that, in spite of the qualitative differences in 
$\sigma(T)$ in the two regimes (with and without local magnetic moments), the 
effect of the parallel $B$ on $\sigma$ is remarkably similar.  This strongly
suggests that in our system {\em at low fields} ($B<2$~T), the critical fields
$B_c\sim B_{\sigma}$, where $B_{\sigma}$ is the polarization field.  Indeed,
in that regime $B_c$ in Fig.~\ref{phase} is comparable to $B_{\sigma}$ in 
Ref.\cite{sergey_scaling} for the same $n_s$.  In addition, since
$B_{\sigma}\propto\delta_n$ continues to hold over a wide range of $n_s$ (up 
to $\delta_n\sim 10$)~\cite{sergey_scaling}, this implies that for 
$\delta_n\sim (0.5-0.6)$ in Fig.~\ref{phase}, the 2D electrons will be fully
spin polarized at $\sim 2$~T.  Therefore, at $B>2$~T the MIT will occur 
between a {\em spin-polarized metal} and a spin-polarized insulator.  This 
conclusion is supported by our observation that $n_c$ no longer seems to 
depend on the magnetic field for $B>2$~T. 

\begin{figure}[t]
\epsfxsize=3.2in \epsfbox{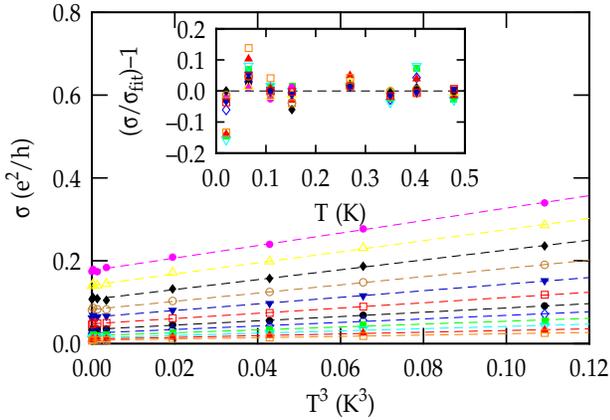}\vspace{5pt}	
\caption{Sample 19, $B=18$~T: $\sigma(T)$ plotted {\it vs.} $T^3$ for 
$n_s(10^{11}$cm$^{-2})$ from 4.9 (top) to 3.8 (bottom) in steps of 0.1; 
$0.020\protect\lesssim T\leq 0.478$~K.  The dashed lines are fits 
$\sigma_{fit}(n_s,T)$.  Inset: The deviation of the data from the fit {\it
vs.} temperature.
\label{tempdep}}
\end{figure}

The detailed properties of the apparent spin-polarized 2D metallic phase and 
the concomitant MIT require further careful investigation that is beyond the 
scope of this Letter.  Here we demonstrate only that the metallic phase 
persists
even in fields as high as 18~T, where we estimate~\cite{sergey_scaling} that 
the 2D system is fully spin polarized in the range of $n_s$ studied.  We find 
that $\sigma (T)$ in the metallic phase follows a simple power-law form
\begin{equation}
\sigma(n_s,B,T) = \sigma(n_s,B,T=0) + C(n_s,B)\,T^{\alpha(B)},
\label{nfit}
\end{equation}
similar to the $B=0$ case, where $\alpha(B=0)=2.0\pm 0.1$~\cite{novel}.  In 
both cases, finite values of $\sigma(n_s,B,T=0)$ indicate that the 2D system 
is in the metallic state.  
At $B=18$~T, however,  $\sigma(T)$ is best described with $\alpha=3$ in 
Eq.~(\ref{nfit}), as shown in Fig.~\ref{tempdep}.  While the mechanism that
leads to $\alpha=3$ is unclear, it is interesting that the $T^3$ term 
(albeit with a different sign) has been observed in the quasi-2D metallic 
Sr$_3$Ru$_2$O$_7$ below 0.5~K near a metamagnetic quantum critical 
point~\cite{grigera}.  In our case, such a jump in magnetization would occur 
as 2D electrons rearrange from an unpolarized to a spin-polarized metallic 
state with decreasing $n_s$ at a fixed high $B$.  The $T^3$ 
variation of resistivity has been also observed in a low carrier density 
metallic state of SrB$_6$~\cite{fisk} but the microscopic mechanism is still 
not understood.

In summary, we have studied the effect of a parallel $B$ on the conductivity of
a 2D electron system with $d\sigma/dT>0$ in the metallic phase, even at $B=0$.
The zero-temperature $(\delta_n,B)$ phase diagram has been constructed based 
on the simple form of $\sigma(T)$ in the metallic regime.  Furthermore, the 
low-field magnetoconductivity has been found to scale with $n_s$, consistent 
with the $(\delta_n,B,T=0)$ phase diagram.  These findings provide strong 
experimental evidence for a quantum phase transition in both $B=0$ and $B\neq 
0$.  The metallic phase is found to persist in fields as high as 18~T.  Our 
results strongly suggest that the 2D metallic phase may exist even in the
absence of spin degrees of freedom.

We are grateful to T. Murphy, E. Palm, and K. Walther for technical 
assistance.  This work was supported by an NHMFL In-House Research Program 
grant, NSF Grants DMR-9796339 and DMR-0071668, NHMFL through NSF Cooperative 
Agreement DMR-9527035, and by ARO-EL-35059.


\vspace{-8pt}

\begin{references}
\vspace{-36pt}
%
\bibitem[]{feng_add} $^{\ast}$Present address: Terayon Communication Systems,
Santa Clara, CA 95054.

\bibitem{Ab_review} See E. Abrahams, S. V. Kravchenko, and M. P. Sarachik, 
Rev. Mod. Phys. {\bf 73}, 251 (2001), and references therein.

\bibitem{novel} X. G. Feng {\em et al.},
Phys. Rev. Lett.~{\bf 86}, 2625 (2001).

\bibitem{Btempdep} D. Simonian {\em et al.},
Phys. Rev. Lett.~{\bf 79}, 2304 (1997); V. M. Pudalov {\em et al.},
JETP Lett.~{\bf 65}, 932 (1997); M. Y. Simmons {\em et al.},
Phys. Rev. Lett.~{\bf 80}, 1292 (1998); K. M. Mertes {\em et al.},
Phys. Rev. B {\bf 63}, 041101 (R) (2001).

\bibitem{okamoto} T. Okamoto {\em et al.}, Phys. Rev. Lett.~{\bf 82}, 3875 
(1999).

\bibitem{yoon} J. Yoon {\em et al.},
Phys. Rev. Lett.~{\bf 84}, 4421 (2000).

\bibitem{spinpol_vit} S. A. Vitkalov {\em et al.},
Phys. Rev. Lett. {\bf 85}, 2164 (2000).

\bibitem{spinpol} E. Tutuc {\em et al.},
Phys. Rev. Lett. {\bf 86}, 2858 (2001).

\bibitem{beta_Pudalov} V. M. Pudalov {\em et al.},
cond-mat/0103087 (2001).

\bibitem{beta} V. T. Dolgopolov {\em et al.},
JETP Lett. {\bf 55}, 733 (1992); M. R. Sakr {\em et al.},
Phys. Rev. B {\bf 65}, 041303(R), (2001); A. A. Shashkin {\em et al.}, 
Phys. Rev. Lett. {\bf 87}, 266402 (2001).

\bibitem{scale} D. Belitz {\em et al.}, 
Rev. Mod. Phys.~{\bf 66}, 261 (1994).

\bibitem{Lee} P. A. Lee and T. V. Ramakrishnan, Rev. Mod. Phys. {\bf 57}, 287
(1985).

\bibitem{tails} Usually, this is a tail of the $E_{0'}$ subband -- the lowest
subband associated with the four conduction band valleys with the light 
effective mass (0.190~$m_e$) perpendicular to the interface.  See also 
Ref.~\cite{AFS}.

\bibitem{AFS} See T. Ando, A. B. Fowler, and F. Stern, Rev. Mod. Phys.~{\bf 
54}, 437 (1982), and references therein.

\bibitem{DP_MIT} D. Popovi\'{c} {\em et al.},
Phys. Rev. Lett.~{\bf 79}, 1543 (1997).

\bibitem{spinflip} X. G. Feng {\em et al.},
Phys. Rev. Lett.~{\bf 83}, 368 (1999).

\bibitem{subbands} A. B. Fowler, Phys. Rev. Lett.~{\bf 34}, 15 (1975); A. 
Kastalsky and F. F. Fang, Surf. Sci.~{\bf 113}, 153 (1982); U. Kunze, J. Phys.
C {\bf 17}, 5677 (1984); S. Kawaji and N. Nagashima, Surf. Sci.~{\bf 196}, 316
(1988); D. Popovi\'{c}, F. F. Fang, and P. J. Stiles, Solid State Commun.~{\bf
68}, 25 (1988).

\bibitem{moments_comment} For brevity, here we refer to this type of 
scattering as ``local magnetic moments''.

\bibitem{Goldenfeld} N. Goldenfeld, {\em Lectures on Phase Transitions and the
Renormalization Group} (Addison-Wesley, Reading, 1992).

\bibitem{3Dbeta} T. F. Rosenbaum {\it et al.},
Europhys. Lett. {\bf 10}, 269 (1989); S. Bogdanovich {\it et al.},
Phys. Rev. B {\bf 55}, 4215 (1997); M. P. Sarachik {\it et al.},
Phys. Rev. B {\bf 58}, 6692 (1998); M. Watanabe {\it et al.},
Phys. Rev. B {\bf 60}, 15817 (1999).

\bibitem{eng_conf} K. Eng {\em et al.}, in {\em Proceedings of the 25th 
International Conference on the Physics of Semiconductors}, edited by N. Miura
and T. Ando, Springer Proceedings in Physics Vol. 87 (Springer, Berlin, 2001),
p. 741; cond-mat/0012063.

\bibitem{city_scaling} S. A. Vitkalov {\it et al.},
Phys. Rev. Lett. {\bf 87}, 086401 (2001).

\bibitem{sergey_scaling} A. A. Shashkin {\it et al.},
Phys. Rev. Lett. {\bf 87}, 086801 (2001).

\bibitem{grigera} S. A. Grigera {\it et al.},
Science {\bf 294}, 329 (2001).

\bibitem{fisk} H. R. Ott {\it et al.},
Z. Phys. B {\bf 102}, 337 (1997).

\end{references}
\end{document}